\documentclass[oneside,letterpaper,11pt]{article}
\pdfoutput=1

\usepackage[letterpaper,total={6.5in, 9in},top=1in]{geometry}

\usepackage[plain]{fancyref}
\usepackage{xcolor}
\usepackage{hyperref}
\hypersetup{
    colorlinks=true,
    linkcolor=blue,
    filecolor=red,      
    urlcolor=magenta,
    breaklinks=true,
}
\usepackage{url}
\usepackage{booktabs}
\usepackage{paralist}
\usepackage{graphicx}
\usepackage{etoolbox}
\usepackage{flushend}
\usepackage{xspace}

\begin{document}

\newcommand{\ie}{i.e.,\xspace}
\newcommand{\eg}{e.g.,\xspace}
\newcommand{\etal}{et al.\xspace}
\newcommand{\editHere}[1][]{\color{blue} \textbf{Continue Editing From Here: #1:}}
\newcommand{\todo}[2][]{{\color{red} \textbf{TODO:#1:} {#2}}}
\newcommand{\note}[2][]{{\color{purple} \textbf{Note:#1:} {#2}}}
\newcommand{\confirm}[1][]{{\color{orange} {#1} \textbf{#1 Confirm?}}}
\newcommand{\yes}[1]{{\color{green} Yes {#1}}}
\newcommand{\no}[1]{{\color{brown} No {#1}}}

\newcommand{\anonymize}[1]{
  \ifAnonymized {\color{red}{ANONYMIZED}}
  \else #1 
  \fi
}
\newif\ifAnonymized

\newcommand*{\fancyrefrqlabelprefix}{rq}
\frefformat{plain}{\fancyrefrqlabelprefix}{RQ#1}
\Frefformat{plain}{\fancyrefrqlabelprefix}{RQ#1}

\title{Why do Users Kill HPC Jobs?}

\author{Venkatesh-Prasad Ranganath \hspace{1cm} Daniel Andresen \\
Kansas State University, USA\\
\{rvprasad,dan\}@k-state.edu
}

\maketitle

\begin{abstract}
Given the cost of HPC clusters, making best use of them is crucial to improve infrastructure ROI.  Likewise, reducing failed HPC jobs and related waste in terms of user wait times is crucial to improve HPC user productivity (aka human ROI).  While most efforts (\eg debugging HPC programs) explore technical aspects to improve ROI of HPC clusters, we hypothesize non-technical (human) aspects are worth exploring to make non-trivial ROI gains; specifically, understanding non-technical aspects and how they contribute to the failure of HPC jobs.

In this regard, we conducted a case study in the context of Beocat cluster at Kansas State University.  The purpose of the study was to learn the reasons why users terminate jobs and to quantify wasted computations in such jobs in terms of system utilization and user wait time.  The data from the case study helped identify interesting and actionable reasons why users terminate HPC jobs.  It also helped confirm that user terminated jobs may be associated with non-trivial amount of wasted computation, which if reduced can help improve the ROI of HPC clusters. 
\end{abstract}

\sloppy
\section{Motivation}
\label{sec:motivation}

Given the cost of creating and operating high-performance computing (HPC) clusters, making best use of the clusters is crucial for \emph{infrastructure ROI}, \eg creation of a level 3 XSEDE \cite{XSEDE:URL} cluster like Beocat at Kansas State University (described in \Fref{sec:study-cluster}), can easily costs more than 2 million US dollars.  Beyond merely keeping processors busy and memory/storage occupied, this is about the usefulness of computations performed on clusters, \ie the results from computations are not wasted due to them being incomplete or incorrect or irrelevant.

This latter goal is often pursued by exploring techniques to reduce HPC job failures stemming from hardware and/or software failures.  In particular, there has been considerable interest in the HPC community to improve infrastructure ROI by identifying and tackling hurdles rooted in technical aspects of HPC.  For example, in 2017, DOE published a technical report focused on needs and ways to specify, test/verify, and debug massively parallel programs \cite{DOE-HPC-Correctness:TR2017}.  There have been empirical studies to characterize and understand job failures by considering 1) various non-human factors such as spatial and temporal dependences between failures, power quality, temperature, and radiation \cite{ElSayed:DSN2013}, and 2) different statistics such as mean time between failures, mean time to repair \cite{Schroeder:IEETDSC10}, and submission inter-arrival time \cite{Yuan:CMA12}.  Ahrens \etal\ studied the use of HPC for data-intensive science in the US DOE and identified various challenges: support to monitor progress of computation and steer computation in real time, use novel and apt data abstractions and representations, and leverage couplings between experiments \cite{Ahrens:CSE11}.  Faulk \etal\ attempted to define and measure HPC productivity in terms of science accomplished and the involved artifacts \cite{Faulk:HPCA04}.

While improving infrastructure ROI is important, we conjecture the community should also focus on improving human ROI.  By \emph{human ROI (aka HPC user productivity)}, we mean 
\begin{itemize}
  \item the effort expended by users to use HPC clusters, \eg writing programs, configuring programs, setting up data, deploying programs, debugging failures, fixing programs and configurations, exploring computational alternatives (\ie alternative data structures and algorithms), and 
  \item the gains and losses incurred by users while using HPC clusters, \ie speedy completion of large computations, wait times for job completion, wait and computation time from failed jobs.
\end{itemize}

Since software is a key component of HPC, there have been numerous efforts exploring the interaction between software engineering and HPC from as early as 2001.  These efforts have attempted to understand if and how various software engineering aspects influence development and use of scientific software, \eg architecture, platforms, programming model, tools/IDEs, effort estimation, developer experience, user preferences, code complexity, portability, performance \cite{Basili:IEEESoftware08,Pancake:PC01,Hochstein:SC05,Shull:ISESE05,Hochstein:CTWatchQuarterly06,Carver:ICSE07,Wienke:SC16}.  They have also attempted to validate conjectures about user productivity in HPC community by interviewing HPC users \cite{Wolter:CTWatchQuarterly06}.  Moving beyond software, McCracken \etal\ have explored HPC workflows to devise better methods to model and measure HPC productivity by considering bottlenecks in HPC user work flows \cite{McCracken:SEHPC07}.  Unlike most efforts related to infrastructure ROI, these efforts are focused on improving HPC user productivity by identifying and alleviating user pain points in using HPC --- the human aspect of HPC.

Similar to these efforts, we believe human aspects should be studied and understood to improve both infrastructure ROI and human ROI.  Possible issues rooted in or related to human aspects are limited awareness about used HPC apps, bad estimations of resources (\eg processors/compute, memory, storage, and runtime) required by jobs, combination of incorrect versions of apps and libraries, and user's interaction with the infrastructure via bad (unintuitive) interfaces.  By understanding such issues, HPC community can implement changes to workflow, interfaces, training, and possibly infrastructure to alleviate such issues and improve human ROI and infrastructure ROI.

In this context, we are interested in \emph{wasted computations: computations whose results are lost or discarded.}  Such computations stem from jobs that are terminated by 

\begin{enumerate}
  \item \textit{the system}, \eg due to a crashing file system, 
  \item \textit{the scheduler}, \eg due to exceeding requested runtime, or 
  \item \textit{the user}, \eg due to non-convergence of computation.  
\end{enumerate}

Independent of the reasons for such terminations, the resources (\eg compute (CPU) time, user wait (wall clock) time) consumed by such terminated jobs are lost and cannot be recovered.  So, learning about such jobs and the reasons for their termination can help quantify, characterize, and prevent wastage of resources and improve system utilization and user productivity. 

Information about when, how, and why a job was terminated in cases 1 and 2 (listed above) is often available in various logs and can be used post hoc to identify reasons for the termination.  Information about users' reasons for terminating jobs in case 3 is seldom recorded and available for post hoc analysis.  Also, the extent of resource wastage in case 3 is often unclear.  Further, most existing efforts focus on cases 1 and 2 by exploring logs to understand job failures.   Consequently, we decided to focus on case 3.  \emph{We conducted a case study of user terminated jobs on Beocat cluster at Kansas State University (KSU).  The purpose of this study was to learn the reasons why HPC users terminate jobs and quantify the wastage from such jobs.}

In the rest of this paper, we will state the questions we want to answer with this study and describe our approach (including design choices) to collect the required data.  We will then answer the posed questions using the collected data.  We will also share our observations and recommendations from the study.  Finally, we will list ways to extend this effort in the future.

\section{Case Study}
\label{sec:study}

\subsection{Questions}
\label{sec:study-questions}

Before we started the case study, we identified the following questions to answer about case (3).  At the same time, we identified the information required to answer these questions.

\begin{enumerate}[{RQ}1]
  \item \emph{What are the reasons for users to terminate jobs?}  To answer this question, when a job was terminated by a user, we need to contact the user and gather the reason for deleting the job.
  \item \emph{How often do users terminate jobs?}  To answer this question, we need to identify jobs that were terminated by users (as opposed to by the system of the scheduler).
  \item \label{rq:3} \emph{How much compute resource is wasted due to user terminated jobs?}  This question attempts to quantify wasted computation in terms of consumed compute resources.  To answer this question, we need to gather various runtime statistics about user terminated jobs, \eg runtime.
  \item \label{rq:4} \emph{How do user terminated jobs compare to system and scheduler terminated jobs (cases 1 and 2) and all jobs executed on the cluster in terms of consumed compute resources?}  This question explores the significance of computation wasted in case 3 relative to computation wasted in cases 1 and 2 and overall system utilization.  This question can be answered using the data collected to answer \fref{rq:3}.
  \item \label{rq:5} \emph{How does wasted computation translate into user wait times, \ie the wall clock time taken by a job to execute?}  This question attempts to quantify wasted computation in terms of user productivity.  To answer this question, we need to gather various scheduling statistics about user terminated jobs, \eg submission time, completion time.
  \item \emph{How do user terminated jobs compare to system and scheduler terminated jobs (cases 1 and 2) and all jobs executed on the cluster in terms of user wait times?}  This question is similar to \fref{rq:4} but focuses on user productivity.  It can be answered using the data collected to answer \fref{rq:5}.
\end{enumerate}

\subsection{HPC Cluster}
\label{sec:study-cluster}

We conducted our study on \emph{Beocat}, the KSU research computing cluster \cite{Beocat:URL}.  It is currently the largest academic supercomputer in Kansas.  Its hardware includes nearly 400 researcher-funded computers, approximately 2.8PB of storage and 7,900 processor cores on machines ranging from dual-processor Xeon e5 nodes with 128GB RAM with 100GbE to six 80-core Xeons with 1TB RAM connected by 40-100Gbps networks.

It utilizes the `condo' model, where individual research groups purchase compute nodes and have priority on them, and other users' jobs on those nodes are killed if required to satisfy the owners' needs.  It is supported by various NSF and university grants, and it acts as the central computing resource for multiple departments across campus.  It is used in classes that cover topics like bioinformatics, big data, cybersecurity, economics, and chemistry.  

Scheduling systems for HPC clusters include software packages such as Slurm, Torque/Maui, PBS, and Sun Grid Engine (SGE). Each of these systems allow users to submit jobs to the job queue, then schedule the jobs based on system-defined priorities to match the stated needs of the job, which are provided by the user, to system resources. The schedulers attempt to balance high performance (maximizing the number of calculations per second for individual jobs), high throughput (maximizing the jobs completed per unit time), and responsiveness (minimizing job wait times and turnaround times for individual jobs). They generally are configured to take advantage of jobs that are killed or end earlier than their alloted time via 'backfilling' queued jobs into the system. 

Beocat staff includes two full-time system administrators with over ten years experience in high-performance computing, a full-time application scientist with a PhD in Physics and 25 years experience optimizing parallel programs and assisting researchers, and part-time director.

Helping users get most out of HPC has historically taken two primary paths: user education and efficient schedulers.  Cyberinfrastructure (CI) facilitation has taken the form of local expertise, local training, personal networks, and more national efforts like the XSEDE Campus Champions, Software Carpentry, and online training/courses such as ``Supercomputing in Plain English''.  Together, these efforts provide a spectrum of assistance and training to users.  At KSU, we have taken advantage of virtually all of these efforts.

\subsection{Design}
\label{sec:study-design}

\subsubsection{Data Collection}
\label{sec:study-data-collection}

During the study period, Beocat used the open-source version of Sun Grid Engine (SGE) to schedule and manage jobs \cite{SGE:URL}.  SGE provides a set of command-line tools to manage jobs such as \texttt{qdel} to terminate jobs.  For data collection, we intercepted user's action to terminate jobs with a custom version of \texttt{qdel} and collected the reasons for job terminations along with some execution data (\eg runtime) reported by SGE tools.

SGE maintains an accounting file containing information about submitted jobs such as job id, user id, submission time, execution start time, execution end time, exit status, cpu time usage, and max memory usage \cite{SGE-Accounting:URL}.  To gather runtime information about jobs, we considered entries in this file for jobs submitted during the study period.

\subsubsection{Design Choices}
\label{sec:study-design-choices}

Since Beocat is "collectively" owned by different research groups, we could not mandate Beocat users to provide reasons for every job termination.  Instead, we could only request and facilitate users to provide reasons for terminating jobs.  To accomplish this, we experimented with two options.

\begin{itemize}
  \item \textbf{Offline Option} When a user terminated a job via \texttt{qdel}, our (first set of) customizations to \texttt{qdel} captured user id and job id, constructed a link to a Google form partially filled with captured information, and emailed this link to the user.  At a later time, the user could fill out the form with the reasons for terminating the job, the scientific application used in the job, and other comments about the job.  If the user did not fill out the form, then we did not have any record of user deleting the corresponding job.
  \item \textbf{Online Option} When a user terminated a job via \texttt{qdel}, our (second set of) customizations to \texttt{qdel} allowed the users to provide the reasons for terminating the job and the names of the scientific applications used in the job via optional command-line options when terminating jobs.  In addition, the customizations captured more detailed information about the job from the SGE accounting file.  When the user did not use the command-line options to provide reasons for terminating the job, we fell back to the offline option to collect reasons via the Google form. 
\end{itemize}

In both options, there were no predefined set of reasons that users could pick from as reasons for terminating jobs.  Instead, users stated their reasons in their own words.

\subsection{Execution}
\label{sec:study-execution}

We conducted our study from \emph{August 15, 2016} to \emph{December 31, 2017}, both dates inclusive.  The offline option was used for the entire study period.  The online option was deployed on \emph{March 5, 2017} and used for the rest of the study period.  

In terms of intervention, we informed Beocat users about our effort (via Beocat's mailing list) and requested them to provide reasons for terminating jobs.  Also, in September 2017, the director of Beocat reached out to the lead of each research group using Beocat and urged their group members to provide input for the study.

\section{Findings}
\label{sec:findings}

In the study period, based on SGE accounting data, 649,542 jobs were submitted to Beocat by 334 unique users (students and researchers) from different departments at KSU and possibly other institutions.  Out of the 649,542 jobs, 639,109 jobs were executed by Beocat, \ie they consumed up non-zero amount of CPU time according to SGE accounting data.  

In the rest of this section, we will answer the questions identified in \Fref{sec:study-questions} and describe few observations made in the study.

\subsection{Answers}
\label{sec:findings-answers}

\subsubsection{RQ1}
\label{sec:findings-rq1}

At the end of the study period, we compiled the reasons provided by users for job termination.  Since users were allowed to state the reasons in their own words, the reasons were manually grouped based on their mutual similarity.  The application scientist on the Beocat team performed this grouping because most Beocat users consult with the application scientist and, hence, he is familiar with the kind of jobs run by different users, the kind of errors they encounter, and the manner in which they interpret errors and communicate about errors.  Consequently, the number of errors introduced by this grouping is likely to be low.  To further limit errors, the application scientist grouped the reasons that were too fuzzy or unclear into a dedicated group \textbf{(o)}.  Likewise, jobs terminated without providing reasons were added to a dedicated group \textbf{(u)}.

The id of the groups along with the corresponding users' reasons for terminating jobs are given in \Fref{tab:reason-groups}.  The distribution of various statistics of user terminated jobs across these groups is provided in \Fref{tab:reason-distribution}.  CPU time is the cumulative CPU time of the jobs that were terminated by the users for the reason represented by the group, \ie the amount of time these jobs would have kept a single CPU core busy.  WC time is the cumulative wall clock time of the same jobs, \ie the amount of time a user would have waited for the jobs to complete if they were executed sequentially.\footnote{This view precludes parallel jobs but admits parallelism within jobs (\eg via multi-threading or MPI).}


\begin{table}
  \centering
  \ifdef{\IEEEtransversionmajor}{
    \caption{Reasons for users terminating jobs}
  }{}
  \begin{tabular}{cl}
    \toprule
    Id & Reason \\
    \midrule
    \textbf{a} & Exploring and testing Beocat\\
    \textbf{b} & System errors\\
    \textbf{c} & Incorrect application parameters\\
    \textbf{d} & Decided to change application parameters\\
    \textbf{e} & Computation has converged \\
    \textbf{f} & Computation is not converging\\
    \textbf{g} & Application code crashed or encountered errors\\
    \textbf{h} & Job script encountered errors\\
    \textbf{i} & Decided to change job parameters\\
    \textbf{j} & Issues with requested amount of memory\\
    \textbf{k} & Job will not finish on time\\
    \textbf{l} & Testing or debugging code\\
    \textbf{m} & External user error\\
    \textbf{n} & Conflicts with other submitted jobs\\
    \textbf{o} & \emph{Unable to understand the provided reason}\\
    \textbf{p} & Inefficient use of resources\\
    \textbf{u} & \emph{No reasons were provided} \\
    \bottomrule
  \end{tabular}
  \ifundef{\IEEEtransversionmajor}{
    \caption{Reasons for users terminating jobs}
  }{}
  \label{tab:reason-groups}
\end{table}

\begin{table*}
  \centering
  \ifdef{\IEEEtransversionmajor}{
    \caption{Distribution of user terminated jobs and their CPU and wall clock (WC) times across various reasons for termination}
  }{}
  \begin{tabular}{crrrrrrr}
    \toprule
    Reason & Frequency & Executed & CPU Time (s) & CPU Time \% & WC Time (s) & WC Time \% \\
    \midrule
    a & 45 & 31 & 767,466,970 & \textbf{10.41} & 702,713,085 & \textbf{32.50}\\
    b & 858 & 408 & 744,614,233 & \textbf{10.10} & 131,126,694 & \textbf{6.06}\\
    c & 501 & 245 & 85,611,725 & 1.16 & 23,468,057 & 1.09\\
    d & 91 & 64 & 126,222,549 & 1.71 & 8,671,989 & 0.4\\
    e & 330 & 328 & 368,006,162 & \textbf{4.99} & 34,852,257 & 1.61\\
    f & 229 & 224 & 293,809,823 & \textbf{3.98} & 20,380,233 & 0.94\\
    g & 77 & 45 & 9,263,731 & 0.13 & 26,168,964 & 1.21\\
    h & 2,666 & 1,286 & 92,835,669 & 1.26 & 118,014,801 & \textbf{5.46}\\
    i & 801 & 441 & 135,507,364 & 1.84 & 46,431,290 & 2.15\\
    j & 37 & 19 & 10,987,872 & 0.15 & 832,454 & 0.04\\
    k & 73 & 39 & 227,105,011 & \textbf{3.08} & 113,021,795 & \textbf{5.23}\\
    l & 123 & 63 & 2,406,771 & 0.03 & 107,799 & 0\\
    m & 85 & 80 & 488,141 & 0.01 & 685,076 & 0.03\\
    n & 1,105 & 405 & 367,008,071 & \textbf{4.98} & 45,938,893 & 2.12\\
    o & 1,430 & 663 & 722,317,595 & \textbf{9.79} & 82,761,133 & \textbf{3.83}\\
    p & 60 & 35 & 60,506,573 & 0.82 & 4,268,953 & 0.2\\
    u & 18,456 & 9,222 & 3,360,871,144 & \textbf{45.57} & 802,912,777 & \textbf{37.13}\\
    \midrule
    Total & 26,967 & 13,598 & 7,375,029,412 & 100 & 2,162,356,250 & 100\\
    \bottomrule
  \end{tabular}
  \ifundef{\IEEEtransversionmajor}{
    \caption{Distribution of user terminated jobs and their CPU and wall clock (WC) times across various reasons for termination}
  }{}
  \label{tab:reason-distribution}
\end{table*}

\paragraph{Description of Reasons} 

Reason \textbf{(a)} stems primarily from users exploring and learning to use Beocat.  The associated jobs are often executed by new users at or after a Beocat training workshop or by students attending a class that uses HPC.  

Reason \textbf{(b)} accounts for situations where jobs are terminated as their progress is hampered by system issues in the cluster, \eg insufficient disk space, file system slowdowns, network failures.

Reason \textbf{(c)} corresponds to user errors in specifying inputs to applications used in jobs.  Typically, these inputs are command line options and arguments, configuration files, and data files provided to apps.  While these reasons are most often associated with the use of off-the-shelf applications such as weka and VASP.

Reason \textbf{(d)} accounts for situations where, after submitting the job, the user decides to change the inputs to the applications used in the job, \eg consider a different data file.

Reasons \textbf{(e)} and \textbf{(f)} correspond to situations where the computation being performed in the job has either converged to a desirable state (\eg a desired level of accuracy/sufficiency has been reached) or diverged to a state from which it cannot recover.  These reasons almost always associated with the use of off-the-shelf applications such as NAMD and VASP.

Reasons \textbf{(g)} and \textbf{(h)} stem from errors encountered while running either the application or the job script, respectively.  Reason (g) also covers failures of user created programs/scripts in languages such as Python and R.

Similar to reason (d), reason \textbf{(i)} accounts for situations where a user may decide to change job parameters (\eg number of cores, maximum memory) after submitting a job.

User estimates of the amount of memory required for their jobs can often be inaccurate; they are either over estimate or under estimate.  In some situations, due to dynamic memory constraints, submitted jobs might start thrashing and slow down significantly.  Reason \textbf{(j)} accounts for these situations.

At times, the amount of time taken by a job does not meet the expectation of the user and the job is terminated by the user.  Such situations are described by reason \textbf{(k)}.

Reason \textbf{(l)} accounts for situations where, before submitting a long running job, users test their jobs for various forms of correctness, \eg logic, setup, data format.

Reason \textbf{(m)} corresponds to situations where jobs are terminated due to trivial/accidental user mistakes, \eg accidental job submission, accidentally moving or deleting files when the job is in flight, and not placing data in the appropriate folder.

Reason \textbf{(n)} accounts for situations when users juggle with jobs by submitting multiples jobs with different application configuration or job configurations and then terminating jobs with redundant or useless configurations.

After observing the job run for a while, users may realize the jobs are not making efficient use of requested resources (compute and/or memory).  Reason \textbf{(p)} accounts for such situations.

As mentioned earlier, we may have insufficient information about reasons terminating jobs.  Such job terminations are covered by reasons \textbf{(o)} and \textbf{(u)}.

\paragraph{Contribution to Wasted Computations} Of the above reasons, reason (a) does not contribute to wasted computations because, while the results from the jobs are not used beyond the workshop or the course, the results help users learn about HPC.  Similarly, reason (l) does not contribute to wasted computation as the results from the job help users identify bugs in their jobs and prevent larger waste of resources.  Reason (e) does not contribute to wasted computations as the job arrived at a satisfactory result that will be used by the user.  All other reasons contribute to wasted computation and should be further explored to curb wasteful computation.

\paragraph{Top Reasons Contributing to Wasted Computations} In \Fref{tab:reason-distribution}, top reasons that contribute 90\% of wasted computations in terms of CPU time and in terms of wall clock time (user wait time) are \textbf{highlighted}.\footnote{User wait time accounts only for the time a job executes and not for the time it waits in the scheduler's queue before execution.}  Reasons (a), (b), (k), (o), and (u) are top contributors independent of how the wasted computation is measure.  However, as mentioned earlier, we should not consider reason (a) as contributing wasted computation.  Interestingly, reasons (e), (f), and (n) are top contributors in terms of CPU time but not in terms of user wait time and reason (h) is a top contributor in terms of user wait time but not in terms of CPU time.

\paragraph{Recommendations} In terms of the extent of contributions and impact on ROI, jobs terminated due to reason (u) and (o) together contribute 55\% and 41\% of wasted computation in terms of CPU time and user wait time, respectively.  Hence, we need to figure out ways to gather reasons for all job terminations and ensure the reasons are comprehensible (more on this in \Fref{sec:future-work}).

In terms of improving infrastructure ROI, to reduce 10\% of wasted computation due to reason (b), we should improve cluster reliability and reduce system failures.  This will also help improve human ROI by reducing associated user wait times for job completion.  To alleviate wasted computation (4.98\%) due to reason (n), we should help users identify and use useful configurations.  Automated monitoring techniques should be considered to detect convergence of computations and reduce any unnecessary computations (4.99\%) that may be performed by jobs terminated due to reason (e).  Improving support for early detection and avoidance of divergent computations will help alleviate wasted computation (3.98\%) due to reason (f).  Finally, to alleviate wasted computation in terms of both CPU time (3.08\%) and user wait time (5.23\%) due to reason (k), we should consider techniques to help users accurately estimate the amount of time required for their jobs.

In terms of improving human ROI, we should educate HPC users about the infrastructure software thru which users use HPC clusters.  This will help reduce user wait times due to reason (h).  We should also educate HPC users about the common workflow and best practices of HPC.  Its effect can help reduce user wait times on jobs associated with reason (a) and, eventually, help reduce both wasted computation and user wait times due to other reasons.

\subsubsection{RQ2}
\label{sec:findings-rq2}

Of the 649,542 submitted jobs, 26,967 jobs were terminated by users.  Since these jobs constitute less than 4.15\% of the total number of submitted jobs, we may be tempted to ignore user terminated jobs to improve ROI; however, the CPU and wall clock times suggest otherwise.

\subsubsection{RQ3}
\label{sec:findings-rq3}

Of the 26,967 user terminated jobs, based on the SGE accounting data, only 13,598 jobs were executing when they were terminated by users while the remaining 13,369 user terminated jobs were in the scheduler queue when they were terminated.  Hence, only 13,598 jobs contributed to wasted computation.

Based on SGE accounting data, these 13,598 jobs cumulatively executed for 7,375,029,412 CPU seconds, \ie these jobs would have kept a single CPU core busy for this amount of time.   This was the amount of wasted computation observed in our study.  This computation would have simultaneously kept all of the $\sim$7900 cores in Beocat busy for approximately 11 days!!

\subsubsection{RQ4}
\label{sec:findings-rq4}

From the SGE accounting file, we identified jobs that terminated normally and calculated the cumulative CPU time and the cumulative wall clock time.  We did the same for jobs that terminated abnormally.  The sum of these times from these two sets is the total CPU and wall clock times for the jobs executed during the study period.  This data is available in \Fref{tab:system-time}.

Since user terminated jobs are not allowed to complete, they will likely be terminated by the scheduler by sending a specific signal, \eg SIGTERM.  Hence, these jobs will be accounted as abnormally exiting jobs in the SGE accounting file.

So, the CPU time due to user terminated jobs (Total CPU Time in \Fref{tab:reason-distribution}) constituted 42.25\% of the CPU time consumed by abnormally exiting jobs and 9.56\%of the CPU time consumed by all executed jobs (in \Fref{tab:system-time}).  In million dollar infrastructures, $\sim$10\% wastage is non-trivial. 

\begin{table}
  \centering
  \ifdef{\IEEEtransversionmajor}{
    \caption{Cumulative CPU and wall clock times for all executed jobs during the study period}
  }{}
  \begin{tabular}{crr}
    \toprule
    Normal Exit & CPU Time (s) &  WC Time (s)\\
    \midrule
    Y & 59,664,147,967 & 13,865,524,891\\
    N & 17,452,418,827 & 3,088,336,839\\
    \midrule
    Total & 77,116,566,794 & 16,953,861,730\\
    \bottomrule
  \end{tabular}
  \ifundef{\IEEEtransversionmajor}{
    \caption{Cumulative CPU and wall clock times for all executed jobs during the study period}
  }{}
  \label{tab:system-time}
\end{table}

\subsubsection{RQ5}
\label{sec:findings-rq5}

Based on the SGE accounting data, 13,598 user terminated jobs cumulatively executed for 2,162,356,250 wall clock seconds, \ie one user would have waited this long (a whooping 25,975 days!!) if these jobs were executed sequentially and then terminated.  This was the amount of wasted user wait time observed in our study.

While 25,027 days is not trivial, this number should not be interpreted simplistically as multiple jobs may have executed simultaneously, a single user may executed multiple jobs simultaneously, and multiple users may have used the cluster simultaneously.  Also, users may not have waited and examined the computation of certain jobs before deciding to terminate them \eg jobs terminated for reasons (d).  So, we plan to consider these nuances and analyze the data in future work to understand how this translates into actual waste in terms of user wait time.

\subsubsection{RQ6}
\label{sec:findings-rq6}

The wall clock time (user wait time) due to user terminated jobs (Total WC Time in \Fref{tab:reason-distribution}) constituted 70.02\% of the wall clock time consumed by abnormally exiting jobs and 12.75\% of the wall clock time consumed by all executed jobs (in \Fref{tab:system-time}).

\subsection{Observations}
\label{sec:findings-observations}

\subsubsection{Offline vs Online Data Collection}
\label{sec:offline-online}

Looking back at the two options we used to collect reasons for terminating jobs, we believe the online option works better the offline option.

\begin{figure*}[!t]
  \centering
  \includegraphics[width=0.9\textwidth]{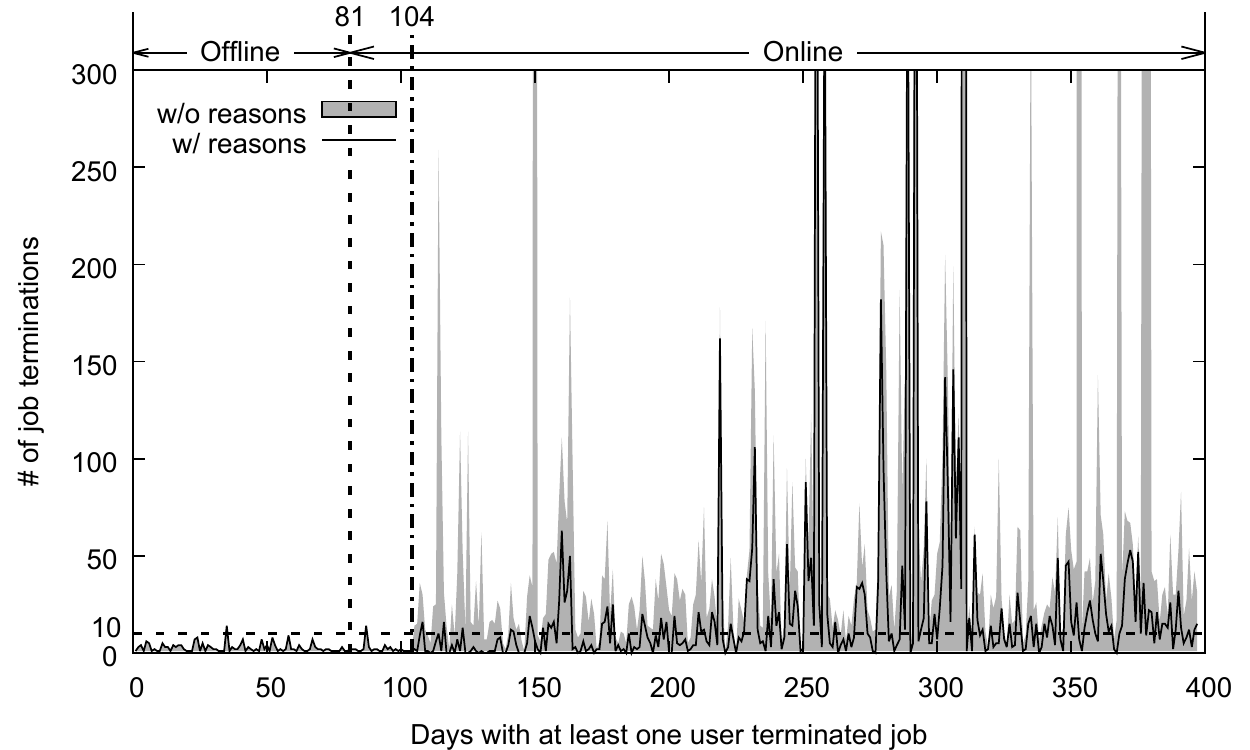}
  \caption{Trend of frequency of user provided reasons}
  \label{fig:reasons-trend}
\end{figure*}

To understand why, consider the trend of frequency of user provided reasons over the study period in \Fref{fig:reasons-trend}.  For each day in the study period with at least one user terminated job as indicated by the collected data, the gray shaded area in the graph shows the number of job terminations and the solid black line shows the number of job terminations with user provided reasons.

When only offline option was deployed from August 15, 2016 thru January 18, 2017, there were 81 days with at least one user terminated job (marked by the vertical dashed line).  During this period, we observed 222 jobs were terminated by users and users provided reasons for each of these job terminations.  While it seemed like we got reasons for every job termination in this period (see Offline part of the graph), this was not the case due to an issue in how we were tracking job terminations.  Instead of tracking job terminations independent of users providing reasons for job terminations, we incorrectly relied on the gathering of reasons for job terminations to track job terminations.

In addition, users provided reasons for very few number of job terminations when only offline option was used; almost always below 10 (marked by the horizontal dashed line at the bottom of the graph).  We conjecture this was due to the disconnect between the act of terminating a job via \texttt{qdel} and the act of providing reasons at a later time via the web form.

Based on these observations, we started testing out the online option from January 18, 2017 and deployed a stable version of the online option on March 5, 2017 (marked by the vertical dotted-dashed line).  After this deployment, the number of observed user terminated jobs increased to 26,204 as every invocation of \texttt{qdel} was used to track job terminations independent of users providing reasons for job terminations (see Online part of the graph).  Also, as the reason for job termination could be provided at the time of terminating a job, we observed users providing reasons for 8,243 job terminations.

Further, we conjecture that more accurate reasons were provided when online option was used as the reasons were provided at the time of terminating jobs.

\subsubsection{Data Quality Issues}
\label{sec:data-quality-issues}

Reflecting back on the case study, we believe there are four issues that affect the data quality.

\begin{itemize}
  \item[\textbf{Missing Reasons}] As observed in \Fref{tab:reason-distribution}, we do not have (useful) reasons for termination of 73.74\% of user terminated jobs (reasons (o) and (u)).  This proportion remains same even when we consider only those user terminated jobs that were executed.  The data suggests that every user does not consistently provide reasons for terminating jobs (based on frequency of reason (o)) and few users consistently fail to provide reasons for terminating jobs (based on frequency of reason (u)).
  \item[\textbf{Ungathered Reasons}] There may have been situations where jobs ran to completion (including crashing) and their results were discarded.  This is possible if the user realizes the utility of the results after the job completes, if the user fails to terminate after realizing it is not needed, or if the job crashes.  Such jobs would not be considered as user terminated; hence, they would incorrectly not have contributed to wasted computation and we would have missed to gather reasons for their reasons to be discarded.
  \item[\textbf{Inconsistent Reasons}] Due to lapse in human memory, users may not provide the same reasons for identical job terminations when the job terminations occur far apart in time.  
  \item[\textbf{Misclassified Reasons}] Due to rater biases and human errors, user provided reasons that should have been classified as identical (\ie put in the same group) may possibly be misclassified as being different in the grouping task described in \Fref{sec:findings-rq1}.
\end{itemize}

These issues need to be addressed to help improve data quality and, subsequently, the findings from such data.  Here are few recommendations to improve data quality.

The number of missing reasons can be reduced by requiring/mandating users to provide reasons when they terminate jobs.  With this approach, to ensure the users provide correct reasons for job terminations, involve users in the effort by 1) periodically sharing the findings with users and, 2) making observable operational changes (based on the findings) to help users.  

To address the issues of inconsistent reasons and misclassified reasons, identify most of the common reasons and allow the users to pick from these reasons.  To help users identify the most applicable reason to a situation, provide real world examples to illustrate the kind of situations described by a specific reason.  Since not all situations may be covered by the set of common reasons, provide users with the option to state their reasons in their own words.

As for ungathered reasons, users to provide brief feedback about every completed job and automatically collect system level data about every job, \eg used applications and libraries and errors encountered during job execution.  The feedback could merely confirm if a job's result was useful and, if not, state the reasons why it wasn't useful; these reasons could be derived from the above set of common reasons.  The system level data can be used to identify crashed jobs and automatically infer likely reasons for their failure.

\subsection{Threats to Validity}
\label{sec:findings-threats-to-validity}

\paragraph{Internal Validity} As seen in \Fref{tab:reason-distribution}, of the 13,598 terminations of jobs that executed for non-zero time, we have reason for 3,713 terminations (27\%).  Since the reasons for 8,201 (9,222 + 663) (73\%) of the terminations is unavailable or unknown, our observations should be considered with caution.  To address this issue, as we continue this effort, we plan on getting user feedback on every job termination.

We have take extreme care to avoid errors while collecting and analyzing data.  Even so, there could be bugs in our study in the form of missed data points, misinterpretation of data, and automation errors.  This threat to validity can be addressed by examining our data set (both raw and processed), analyzing the automation scripts, and repeating the experiment.

\paragraph{External Validity} We conducted our study on Beocat, a level 3 XSEDE cluster \cite{XSEDE:URL}, which may not be representative of other clusters in terms of user base, the collection of installed and used software, or the hardware composition of the cluster.  So, our observations from this study are immediately applicable only to clusters that are comparable Beocat and should be generalized only after replicating the study.

\subsection{Case Study Artifacts}
\label{sec:findings-artifacts}

All of the data and scripts from the case study are hosted in a public Git repository available at \url{https://bitbucket.org/rvprasad/why-do-users-kill-hpc-jobs}.

\section{Future Work}
\label{sec:future-work}

Based on the findings and observations in this study, here are few ways to improve our understanding of how non-technical and human factors can affect ROI of HPC clusters.

\begin{itemize}
  \item Disentangle user wait time data to gain better understanding of the impact of wasted computation on user wait time.
  \item Gather accurate data about reasons why the system terminated a job, \eg memory or run time limits were exceeded, unavailable files/resources were accessed, system was shutdown.
  \item Gather more accurate data when users terminate jobs; specifically, data about reasons that cannot be captured by the system.
  \item Automatically gather system level data when a job terminates, \eg which libraries (including their versions) were loaded as part of a job, what exceptions/errors were reported by the system in the context of a job.
  \item Automatically gather system level data about aspects that may affect jobs, \eg libraries (including their versions) requested by a job, libraries (including their versions) loaded for a job, time when a library was updated on the system, old and new versions of an updated library, record time periods and the system profile when a job was thrashing.
  \item Replicate the study on other clusters (\eg other level 3 XSEDE clusters) that are similar to Beocat and compare the findings from such studies.
  \item Conduct similar studies on other clusters that are not similar to Beocat, \eg level 1 and 3 XSEDE clusters.
\end{itemize}

\section{Summary}
\label{sec:summary}

In this paper, we have described a case study of user terminated jobs on Beocat cluster at Kansas State University.  The purpose of this study was to learn the reasons why users terminate jobs and to quantify wasted computations in such jobs.  Besides identifying a list of reasons why users terminate jobs, the case study suggest that user terminated jobs may be associated with non-trivial amount of wasted computation.  Reducing such wasted computation may be a good opportunity to improve both infrastructure and human ROI of HPC clusters.

We hope this study will spark a conversation in the HPC community to collaboratively explore non-technical (human) aspects to improve ROI for HPC clusters.

\section*{Acknowledgements}
The authors would like to thank Dr. David Turner for creating custom \texttt{qdel} scripts to collect user reasos for job deletions, classifying user provided reasons for job terminations, and providing guidance about the data pertaining to SGE and Beocat.

The authors would also like to thank Adam Tygart for assistance with deploying custom \texttt{qdel} scripts and providing guidance about the data pertaining to SGE and Beocat.


Beocat is supported via a variety of NSF, DOE, NIH, and KSU.

\bibliographystyle{plain}
\bibliography{references} 

\end{document}